\documentclass[aps,amsmath,twocolumn,amssymb,showpacs,floatfix,nofootinbib]{revtex4}
\usepackage{amssymb,amsmath,graphicx,color,microtype}
\usepackage{epsf}
\usepackage{epstopdf}
\usepackage {amssymb}
\newcommand{\nc}{\newcommand}
\nc{\ba}{\begin{eqnarray}}
\nc{\ea}{\end{eqnarray}}
\newcommand\be{\begin{equation}}
\newcommand\ee{\end{equation}}

\nc{\e}{{\bf{e}}}
\nc{\kk}{{\bf{k}}}
\nc{\pp}{{\bf{p}}}

\nc{\bfk}{{\bf{k}}}
\nc{\bfx}{{\bf{x}}}
\nc{\bfp}{{\bf{p}}}

\nc{\eH}{{\epsilon_H}}
\nc{\calP}{{\cal P}}
\nc{\im}{{ \mathrm{Im} } }

\begin{document}
\title{Probing the Running of Primordial Bispectrum and Trispectrum using CMB Spectral Distortions}

\author{Razieh Emami}
\email{razieh.emami$_$meibody@cfa.harvard.edu}
\affiliation{Center for Astrophysics, Harvard-Smithsonian, 60 Garden Street, Cambridge, MA 02138, USA}

\begin{abstract}
We compute the impact of the running of higher order density correlation functions on the two point functions of CMB spectral distortions (SD).  We show that having some levels of running enhances all of the SDs by few orders of magnitude which might make them easier to detect. Taking a reasonable range for $ |n_{f_{NL}} |\lesssim 1.1$ and with $f_{NL} = 5$ we show that for PIXIE like experiment, the signal to noise ratio, $(S/N)_{i}$, enhances to $\lesssim  4000$  and $\lesssim 10$ for $\mu T$ and $yT$ toward the upper limit of $n_{f_{NL}}$. In addition, assuming $ |n_{\tau_{NL}}|< 1$ and $\tau_{NL} = 10^3$, $(S/N)_{i}$ increases to  $\lesssim  8\times 10^{6}$, $\lesssim 10^4$ and $\lesssim 18$ for $\mu\mu$, $\mu y$ and $yy$, respectively. Therefore CMB spectral distortion can be a direct probe of running of higher order correlation functions in the near future.

\end{abstract}

\maketitle
\section{Introduction}
Inflationary universe has proven to be among the most successful paradigms to describe the required seed of the density perturbation. Although most of the single filed inflationary models predict an almost adiabatic, Gaussian and scale-invariant primordial spectrum, there are some scenarios which predict deviation from the above cases. Observational probes are underway to shed more lights about the Physics of the early universe by seeking any deviations from the single field inflation. Planck satellite \cite{Ade:2015ava}  was the most recent CMB experiments to probe the Physics of inflation and it gave us valuable constraints on the amplitude of the non-Gaussianity at the CMB scale, $ 10^{-4} \lesssim k/\rm{Mpc}^{-1} \lesssim 10^{-1} $. These limits are hoped to be more robust using the next generation of the CMB experiments like S4.  There are also many LSS surveys to probe inflation at smaller scales, $ 10^{-1} \lesssim k/\rm{Mpc}^{-1}  \lesssim \mathcal{O}(1) $. However non of them give us any information for scales smaller than \textit{few} $\rm{Mpc}^{-1}$. It was recently mentioned in \cite{Emami:2017fiy} that the limits on the primordial spectrum can be extended to very small scales using the ultra compact minihalo objects (UCMHs) or the Primordial Black Holes (PBHs) which are both based on the astrophysical processes. Here we note another possibility for gaining novel information about the physics of inflation by using the CMB spectral distortion. 
This approach was firstly considered in \cite{Pajer:2012vz, Ganc:2012ae} to probe the scale dependence of the non-Gaussianity with mainly focusing on the correlation of the CMB temperature fluctuations with the chemical potential $(\mu)$ distortion. The approach was further discussed and was extended in \cite{Emami:2015xqa} by adding $y$ distortion in the configuration space. It was also applied as a useful tool to probe different clases of inflationary models \cite{Dimastrogiovanni:2016aul, Cabass:2018jgj,  Haga:2018pdl, Bae:2017tll, Cabass:2017kho}, higher order terms in cosmological perturbation \cite{Ota:2018zwm, Shiraishi:2016hjd, Ota:2016mqd} or evolution of CMB spectral distortion \cite{Chluba:2016aln}. Connecting further to observations, it is also required to seek for any possible late time cross-correlation between the Temperature and the Compton $y$ distortion as was done in  \cite{Creque-Sarbinowski:2016wue,  Ravenni:2017lgw} and to extract foreground obscured in constraining inflation \cite{Remazeilles:2018kqd}.

In an attempt to extend the current theoretical discussions in this context, here we consider the full analysis of $\mu$ and $y$ spectral distortions in the Fourier space by allowing an scale dependent function for the bispectrum and trispectrum. 
This opens up the possibility of probing any shapes of bispectrum and trispectrum of the curvature perturbation. In order to do the analysis analytically, though, we consider a simple form with the power-low scale dependence for both of them. Using this simple form, we estimate the impact of running of the non-Gaussianity and trispectrum in various S/N. 

We include the impact of the thermal Sunyaev Zel'dovich effect in estimating the noise for any quantities that contain the parameter $y$ in the signal to noise analysis. 

The paper is unfolded as the following. In Sec. \ref{mu-And-y} we present a short summary of the spectral distortion from the damping of the acoustic waves in the photon-baryon plasma. 
Then in Sec. \ref{Two-point-correlations} we compute the cross-correlation of CMB temperature with both of $\mu$ and $y$ distortion as well as the auto-correlation of SD parameters with each other. We explicitly show how they are related to the bispectrum and trispectrum of the primordial curvature perturbations. 
In Sec. \ref{detectability} we present the S/N ratio as a systematic way to probe these signals. We conclude in Sec. \ref{conclusion}. 
\section{CMB distortions}
\label{mu-And-y}
In this section we compute the chemical potential $\mu$ and Compton $y$ distortions. Using the combination of Bose-Einstein distribution and the conservation of the total number of photons we calculate SD parameters in terms of the released energy into the plasma as,
\ba
\label{mu}
\mu &\simeq& 1.4 \frac{\delta E}{E} = -1.4 \int_{z_{\mu,i}}^{z_{\mu,f}} \frac{d}{dz} \frac{Q(z)}{\rho_{\gamma}},\\
\label{y}
y &\simeq& \frac{\delta E}{4E} = - \frac{1}{4}\int_{z_{y,i}}^{z_{y,f}} \frac{d}{dz} \frac{Q(z)}{\rho_{\gamma}}.
\ea
where $Q(z) = \rho_{\gamma}(z) \langle \delta^2_{\gamma}(x)\rangle_{p} \frac{c_{s}^2}{\left(1+c_{s}^2\right)}$ refers to the energy injection due to the damping of the acoustic waves in the plasma.  Here $\rho_{\gamma}(z) $ refers to the energy density of photon-baryon plasma and $c_{s} \simeq 1/3 $ denotes the sound speed inside the baryon-photon plasma. As for $ \langle \delta^2_{\gamma}(x)\rangle_{p}$ we use Eq. (5) in \cite{Pajer:2012vz}. Combining all of these factors and going to the Fourier space we get the following expression for $\mu$ and $y$ distortions, 
\ba
\label{mu3}
\mu &=& 4.57  \int \frac{d^3k_1 d^3k_2}{(2\pi)^6} \mathcal{R}(k_1) \mathcal{R}(k_2) \langle \cos{(k_1r)} \cos{(k_2r)}\rangle_{p} \nonumber\\
&& \times ~ \e^{i\vec{k}_+\cdot \vec{x}} W(\frac{k_+}{k_s}) \bigg{[} 
e^{-\frac{(k_1^2 + k_2^2)}{k_D^2}} \bigg{]}_{z_{\mu,f}}^{z_{\mu,i}},\\
\label{y3}
y &=& 0.82  \int \frac{d^3k_1 d^3k_2}{(2\pi)^6} \mathcal{R}(k_1) \mathcal{R}(k_2) \langle \cos{(k_1r)} \cos{(k_2r)}\rangle_{p} \nonumber\\
&& \times ~ \e^{i\vec{k}_+\cdot \vec{x}} W(\frac{k_+}{k_s}) \bigg{[} 
e^{-\frac{(k_1^2 + k_2^2)}{k_D^2}} \bigg{]}_{z_{y,f}}^{z_{y,i}}.
\ea
where $\vec{k}_{+} \equiv \vec{k}_1 + \vec{k}_2$ with $k_{D}$ referring to the diffusion damping scale as was computed in Eq. (3) of \cite{Pajer:2012vz}.

Finally using Eqs. (\ref{mu3}), (\ref{y3}) we compute the ensemble averaged value of the distortions. 
\ba
\label{mu ensemble average}
\langle \mu(x)\rangle &\simeq& 2.28 \int d\log{k} \Delta^2_{\mathcal{R}}(k) \bigg{[} e^{-2k^2 / k_{D}^2} \bigg{]}_{\mu_f}^{\mu_i}, \\
\label{y ensemble average}
\langle y(x)\rangle &\simeq& 0.41 \int d\log{k} \Delta^2_{\mathcal{R}}(k) \bigg{[} e^{-2k^2 / k_{D}^2} \bigg{]}_{y_f}^{y_i}.
\ea
with $\Delta^2_{\mathcal{R}}(k)$ referring to the dimensionless power spectrum of the primordial curvature perturbation,

\ba
\label{power-spectrum}
\langle \mathcal{R}(k_1)\mathcal{R}(k_2)\rangle &\equiv& (2\pi)^3 \delta^3(\vec{k}_{+})P_{\mathcal{R}}(k_1),  \nonumber\\
P_{\mathcal{R}}(k) &\equiv& \frac{2\pi^2 \Delta^2_{\mathcal{R}}(k)}{k^3}.
\ea

\section{Two point correlation Functions of the Temperature and CMB distortions}
\label{Two-point-correlations}
Having presented the general form of the SD parameters here we compute the two point correlation functions of distortions parameters either with the CMB temperature or with each other. It is more  convenient to present these quantities in terms of the spherical Bessel functions as, 
\ba
\label{aT}
a_{lm}^T &\equiv& \int d\hat{n} \frac{\Delta T (\hat{n})}{T} Y^{*}_{lm}(\hat{n}) \nonumber\\
&=& \left(\frac{12}{5} \pi\right) (-i)^l \int \frac{d^3k}{(2\pi)^3} \mathcal{R}(k) \Delta_l(k) Y^{*}_{lm}(\hat{k}).
\ea
where $\Delta_l(k) \simeq j_l(kr_l)/3$, is the Radiation transfer function and we have used the Sachs-Wolfe approximation. Here $r_l$ is the distance from the last scattering surface, $r_l = 14  \rm{Gpc}$ while $j_l$ is the spherical Bessel function. 

In addition we have, 
\ba
\label{amu2}
a_{lm}^\mu &=& (18.3 \pi) (-i)^l \int  \frac{d^3k_1 d^3k_2}{(2\pi)^6} Y^{*}_{lm}(\hat{k}_+)  \mathcal{R}(k_1) \mathcal{R}(k_2)   W(\frac{k_+}{k_s}) \nonumber\\
&&  \langle \cos{(k_1r)} \cos{(k_2r)}\rangle_{p} j_{l}(k_{+} r_L)  \bigg{[}
e^{-\frac{(k_1^2 + k_2^2)}{k_D^2}} \bigg{]}_{z_{\mu,f}}^{z_{\mu,i}}, \\
\label{ay2}
a_{lm}^y &=& (3.3 \pi) (-i)^l \int  \frac{d^3k_1 d^3k_2}{(2\pi)^6} Y^{*}_{lm}(\hat{k}_+)  \mathcal{R}(k_1) \mathcal{R}(k_2) W(\frac{k_+}{k_s})  \nonumber\\
&&   \langle \cos{(k_1r)} \cos{(k_2r)}\rangle_{p} j_{l}(k_{+} r_L) \bigg{[}
e^{-\frac{(k_1^2 + k_2^2)}{k_D^2}} \bigg{]}_{z_{y,f}}^{z_{y,i}}. 
\ea
Using the above forms of $a_{lm}$ we can compute all of the two pints correlation functions. Since the results are proportional to bispectrum and trispectrum of the curvature perturbation, we first present a simple scale dependent model for computing them.  It is interesting to evaluate how does the correlation functions change for different parameters of the model. 

\subsection{Primordial Bispectrum and Trispectrum}
Here we take a simple and generic parametrization for the scale-dependent non-Gaussianity as was presented in \cite{Biagetti:2013sr,Becker:2012je, Byrnes:2010ft}, 
\ba
\label{scale dpendent}
B_{\Phi}(k_1, k_2, k_3) &=& 2 \bigg{[} \xi_{f_{NL}}(k_3) \xi_{m}(k_1) \xi_{m}(k_2) P_{\phi}(k_1) P_{\phi}(k_2) \nonumber\\
&& + 2 perms \bigg{]}.
\ea
where we have defined,
\ba
\label{xi definition}
\xi_{f_{NL}, m}(k) \equiv \xi_{f_{NL}, m}(k_0) \left(\frac{k}{k_0}\right)^{n_{f_{NL},m}}.
\ea
here $\xi_{f_{NL}}$ refers to the strength of bispectrum while $\xi_{m}$ denotes the ratio of the contribution of each field in the matter power spectrum. We choose $k_0 = 0.002 Mpc^{-1}$ as the pivot scale for the CMB. In addition $n_{f_{NL}}$ and $n_{m}$ denotes the running of the bispectrum and power spectrum, respectively. 

In addition, the scale dependent trispectrum is given by,
\ba
\label{trispectrum}
&& T_{\Phi}(k_1, k_2, k_3, k_4) =  \left(\frac{25}{9}\right) \bigg{[} \xi_{\tau_{NL}}(k_3, k_4) \xi_{m}(k_1) \xi_{m}(k_2) \nonumber\\
&& ~~~~~~~~\xi_{m}(k_{13})P_{\phi}(k_1) P_{\phi}(k_2) P_{\phi}(k_{13}) + 11 perms\bigg{]}.
\ea
where  we have $|k_{13}| \equiv |k_1 + k_3|$ and with $\xi_{\tau_{NL}}$, 
\ba
\label{tauNL}
\xi_{\tau_{NL}}(k_i, k_j) \equiv \xi_{\tau_{NL}}(k_0) \left(\frac{k_i k_j}{k_0^2}\right)^{n_{\tau_{NL}}}
\ea
As a comparison, in the ``single field" case we have $\xi_{\tau_{NL}}(k_i, k_j) = \left(\frac{36}{25}\right) \xi_{f_{NL}}(k_i) \xi_{f_{NL}}(k_j)$ and $\xi_{m}(k) =1$. However, in the more generic case, we  have two free parameters, $n_{f_{NL}}$ and $n_{\tau_{NL}}$ which describe the scale dependence of  bispectrum and trispectrum, respectively. We ignore any non trivial scale dependence for the power spectrum which means putting $n_m =0$ in what follows.

We can simplify the above expression using the squeezed limit for the bispectrum,  $k_3 \ll k_1 \simeq k_2$, and the collapsed limit for the trispectrum, $\vec{k}_{12} \simeq 0$. In this limiting case we get, 
\ba
\label{squzeed limit}
\langle \mathcal{R}_{\vec{k}_1} \mathcal{R}_{\vec{k}_2} \mathcal{R}_{\vec{k}_3} \rangle' &=& \left( \frac{12}{5} \right)f_{NL}(k_{-}/2) P(k_{-}/2) P(k_{+}), \nonumber\\
\label{collapsed limit}
\langle \mathcal{R}_{\vec{k}_1} \mathcal{R}_{\vec{k}_2} \mathcal{R}_{\vec{k}_3}  \mathcal{R}_{\vec{k}_4} \rangle' &=& 4  \tau_{NL}(k_{-}/2, k_3) P(k_{-}/2) P(k_{+})\nonumber\\
&& \times P(k_{3}).
\ea
 where $'$ shows that we have removed $(2\pi)^3 \delta^3(\vec{k}_1 + \vec{k}_2 + \vec{k}_3)$. In addition, we have only presented the dominant contributions for both of the correlation functions. which are two terms for the bispectrum at the squeezed limit and four terms for the trispectrum in the collapsed limit. 
 
We also work with the following form for $f_{NL}$ and $\tau_{NL}$,
\ba
\label{scale dependence}
f_{NL}(k) &=& f^{p}_{NL}\left(\frac{k}{k_p}\right)^{n_{f_{NL}}},\\
\tau_{NL}(k_i, k_j) &=& \tau^{p}_{NL}\left(\frac{k_i k_j}{k_p^2}\right)^{n_{\tau_{NL}}}.
\ea
here the index $p$ refers to the pivot  scale. As for the power spectrum, we assume the following sale dependence for the $\Delta^2_{\mathcal{R}}(k)$, 
\ba
\label{scale power}
\Delta^2_{\mathcal{R}}(k) = \Delta^2_{\mathcal{R}}(k_p) \left(\frac{k}{k_p}\right)^{(n_s-1)}.
\ea
\begin{figure*}[t!]
 \center
  \includegraphics[width=\textwidth]{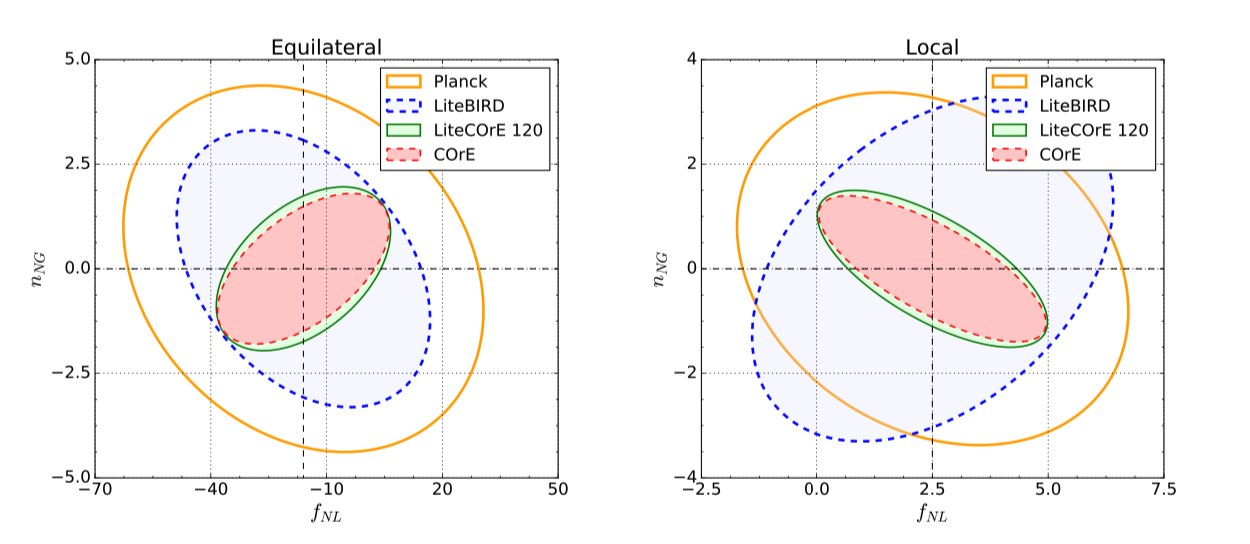}
	\hspace{0.009\textwidth}
	\caption{\label{fnl-forecast}1$\sigma$ error ellipse forecast in the plane of $f_{NL}-n_{f_{NL}}$ for two favorite family of the inflationary models  \cite{Oppizzi:2017nfy}.}
\end{figure*}
\subsection{Cross correlation of the CMB distortion and the Temperature fluctuation}
Here we present the cross correlation of $\mu T$ and $y T$.  For two generic functions $a_{lm}^{x}$ and $a_{lm}^{y}$, the $l$-space two point correlation function is given by,
\ba
\label{correlation function xy}
\langle \left( a_{lm}^x \right)^* a_{l'm'}^y\rangle
= \delta_{ll'} \delta_{mm'} C_{l}^{xy}.
\ea
Using the above expressions, we compute $C_l^{\mu T}$ and $C_l^{y T}$ as,
\ba
\label{muT2}
l(l+1)C_l^{\mu T} & \simeq & 2.2 \pi  \frac{ f_{NL}^p \Delta^4_{\mathcal{R}}(k_p)}{\Gamma_f} \left(\frac{k_D(z)}{\sqrt{2}k_p} \right)^{\Gamma_f}\Bigg{|}_{z_{\mu,f}}^{z_{\mu,i}}, \nonumber\\
l(l+1)C_l^{y T} & \simeq & 0.4 \pi \frac{ f_{NL}^p \Delta^4_{\mathcal{R}}(k_p)}{\Gamma_f} \left(\frac{k_D(z)}{\sqrt{2}k_p} \right)^{\Gamma_f}\Bigg{|}_{z_{y,f}}^{z_{y,i}}. \nonumber\\
\ea
where we have defined $\Gamma_f \equiv \left( n_s -1 + n_{f_{NL}} \right)$. From Eq. (\ref{muT2}) it is clear that the cross correlation of the CMB temperature and distortion parameters is linearly proportional to the value of $f_{NL}$.  In addition, it also depends on the running of the non-Gaussianity.The ratio of the $C_l^{\mu T}$ and $C_l^{y T}$ is computed as,
\ba
\label{ratio-yT- muT}
\left(\frac{C_l^{y T}}{C_l^{\mu T}}\right) = 0.2 \left(\frac{k_{D}(z_{y,i})}{k_{D}(z_{\mu,i})}\right)^{\Gamma_f} .
\ea
It is worth to compare our results with \cite{Emami:2015xqa} where some of us considered the ratio of the $\mu T$ and $y T$ for the simplest possible case of constant $f_{NL}$ and in the configuration space. Here we have generalized this and see the explicit dependency of the above ratio to the running of Non-Gaussianity. 

To proceed, we comment on the possible choices of the running of the non-Gaussianity. In general the value of the running of non-Gaussianity is a free parameters in this formalism.
Theoretically, it can arise from different physical processes like the non-linear evolution of the perturbations, different types of interactions among the fields, modification of sound speed in signle inflation or from the modification of the background metric. It is therefore possible to treat this as a generic prediction of the inflationary models \cite{Oppizzi:2017nfy, Byrnes:2010ft, Takahashi:2014bxa}. Its value can be smaller or order unity depending on the details of the inflationary model as well as the interactions. ( For example Eq. (105) of \cite{Takahashi:2014bxa}). 
We may however constraint its actual value using various observations. To date the best type of forecasts are from the CMB experiments. 
Here we take a realistic choice for this parameter by using the most recent CMB analysis and forecasts as was performed in Ref. \cite{Oppizzi:2017nfy}. In their analysis they have used the WMAP 9 year datas and found bounds ($68\%$ C.L.) for the running of the Non-Gaussianity in few different inflationary models, $ -0.6  <n_{f_{NL}} < 1.4$ for the ``single-filed" curvaton scenario, $ -0.3 <n_{f_{NL}} < 1.2$  for the ``two-field" curvaton models  and finally $ -1.1 <n_{f_{NL}} < 0.7$ for the DBI inflation.  They have also presented a forecasts for $f_{NL}$ vs $n_{f_{NL}}$ using the more recent experimental results. In order to be more clear in Fig. \ref{fnl-forecast} we present the results of their forecasts for $f_{NL}$ vs $n_{f_{NL}}$.  The plot shows a 1$\sigma$ ellipse in the plane of $f_{NL}-n_{f_{NL}}$ for two favorite family of the inflationary models. This is to date the most robust constraints on the value of the running of Non-Gaussianity. 

Here we aim to be as model independent as possible. Therefore we use the smallest possible ellipse in Fig. \ref{fnl-forecast} and we limit ourselves to the following range $ -1.0 <n_{f_{NL}}< 1.5$.  

From the plot we explicitly see that the ratio depends on the running of the non-Gaussianity and it can be bigger than one for negative values of $n_{f_{NL}}$ while get smaller and smaller for positive running. This makes sense as for negative running of non-Gaussianity $\mu$ distortion is smaller than $y$ distortion taking into account that $50  \leq  k_{\mu}/ \rm{Mpc}^{-1} \leq 10^4$ is pushed toward smaller scales as compared with the $1  \leq  k_{y}/ \rm{Mpc}^{-1} \leq 50$.

\begin{figure}[!h]
	\centering
	\includegraphics[width=0.49\textwidth]{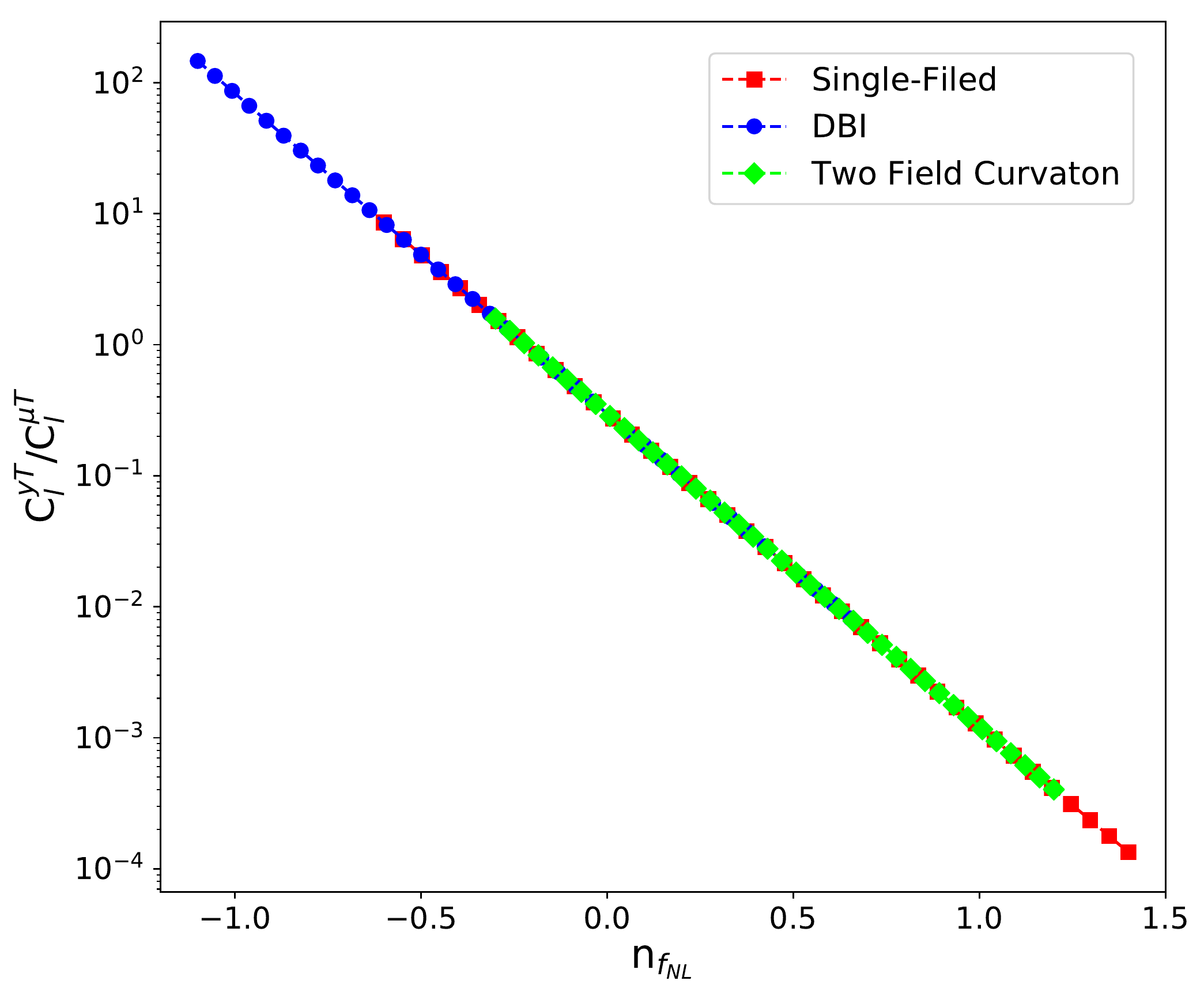}
	\hspace{0.009\textwidth}
	\caption{\label{comparison-mu-y} The behavior of $\left(C_l^{y T}/C_l^{\mu T}\right)$ as a function of $n_{fnl}$ for different ranges of $n_{fnl}$ coming from the CMB experiment.}
\end{figure}

\begin{figure}[!h]
	\centering
	\includegraphics[width=0.49\textwidth]{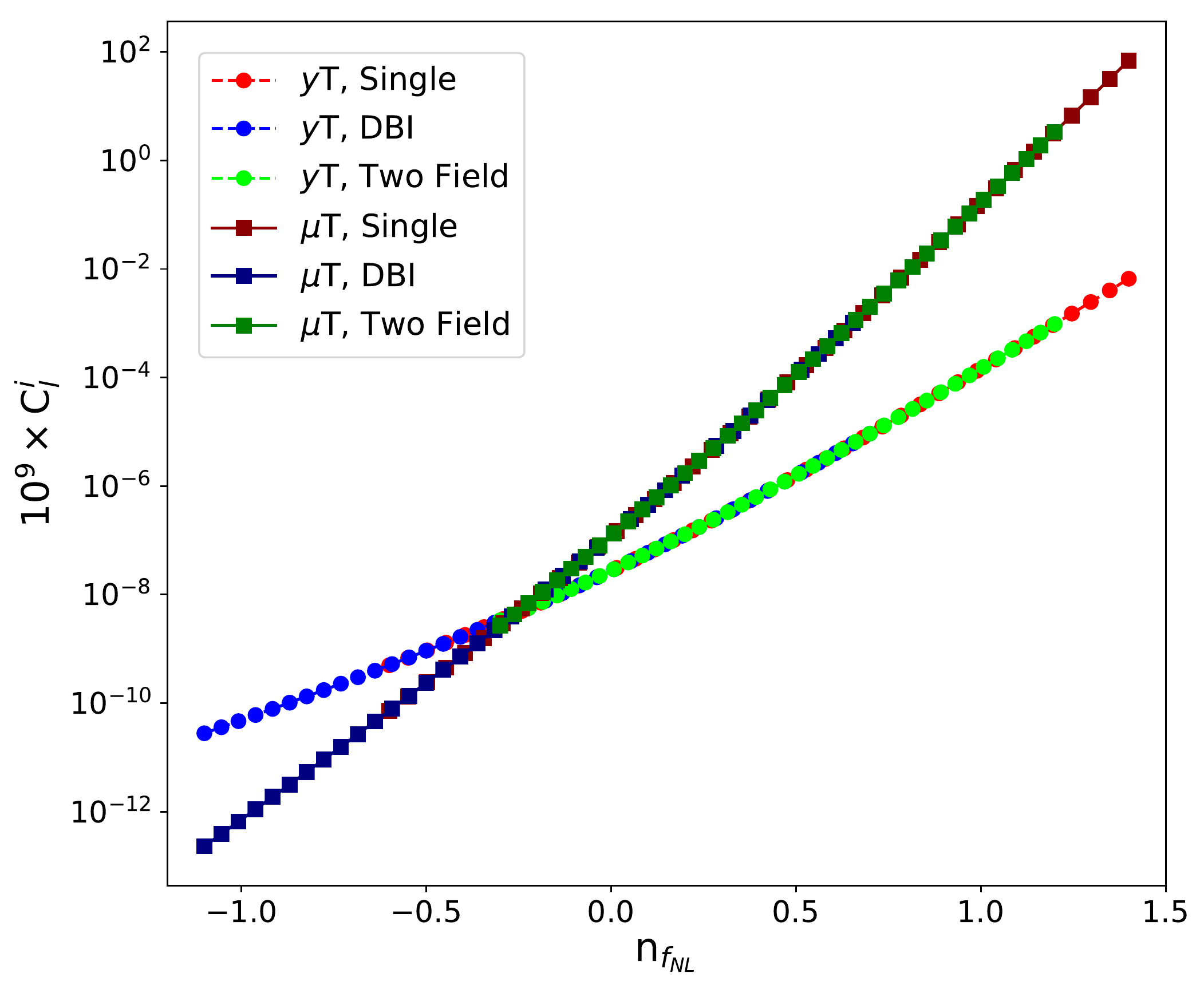}
	\hspace{0.009\textwidth}
	\caption{\label{cl-mu-y} Behavior of $C^i_{l}$ for $ i = (\mu T, y T)$ as a function of $n_{f_{NL}}$ for different ranges of $n_{f_{NL}}$ coming from the CMB experiment. Here we take $f^p_{NL} = 5$.}
\end{figure}

\subsection{Auto/Cross correlations of distortion parameters}
In this section, we compute the two point and cross-correlation of distortion parameters with each other. This includes $\mu \mu$, $\mu y$ and$yy$, 
\ba
\label{mu-mu correlation2}
l(l+1)C_{l}^{\mu \mu} &=& 41.4 \pi \tau^p_{NL} \frac{\Delta^6_{\mathcal{R}}(k_p)}{\Gamma_g^2} \left(\frac{k_{D}(z)}{\sqrt{2}k_p}\right)^{2\Gamma_g}\Bigg{|}_{z_{\mu,f}}^{z_{\mu,i}},\nonumber\\
l(l+1)C_{l}^{\mu y} &=& 7.42 \pi \tau^p_{NL} \frac{\Delta^6_{\mathcal{R}}(k_p)}{\Gamma_g^2} \left(\frac{k_{D}(z)}{\sqrt{2} k_p}\right)^{\Gamma_g}\Bigg{|}_{z_{y,f}}^{z_{y,i}} \nonumber\\
&& ~~\times  \left(\frac{k_{D}(z)}{\sqrt{2} k_p}\right)^{\Gamma_g}\Bigg{|}_{z_{\mu,f}}^{z_{\mu,i}} , \nonumber\\
l(l+1)C_{l}^{yy} &=& 1.33 \pi \tau^p_{NL} \frac{\Delta^6_{\mathcal{R}}(k_p)}{\Gamma_g^2} \left(\frac{k_{D}(z)}{\sqrt{2}k_p}\right)^{2\Gamma_g}\Bigg{|}_{z_{y,f}}^{z_{y,i}}. \nonumber\\
\ea
where we have defined $\Gamma_g \equiv \left( n_{s} -1 + n_{\tau_{NL}}\right)$. The situation is quite similar to  $\mu T$ and $y T$ as we discussed above.

\section{Detectability of the CMB Distortion Anisotropies}
\label{detectability}
Here we estimate the detectability of the CMB distortion with future observations. We start with presenting an estimator for $C_{l}$,
\ba
\label{estimator}
\hat{C}^{XY}_{l} = \left(\frac{1}{2l+1}\right)\sum_{m} a_{lm}^{X} a_{lm}^{Y}.
\ea
It can be shown that under the null hypothesis the variance of this estimator is given by \cite{Miyamoto:2013oua, Knox:1995dq},
\ba
\label{variance2}
\sigma^2_{ll'} &=& \frac{\delta_{ll'}}{2l+1} \left(C_{l}^{XX} + C_{l}^{XX,N} \right) \left(C_{l}^{YY} + C_{l}^{YY,N} \right), \left(X \neq Y\right) \nonumber\\
\label{variance3}
\sigma^2_{ll'} &=& \frac{2}{2l+1} \left(C_{l}^{XX} + C_{l}^{XX,N} \right)^2, \left(X = Y\right).
\ea
where $C_{l}^{XX,N}$ refers to the noise power spectrum of the observation while $C_{l}^{XX}$ stands for the actual signal. 

While the production of $\mu$ distortion stops after $z \simeq 5 \times 10^4$, $y$ distortion can be still created in the late universe through the thermal Sunyaev-Zeldovich (tSZ) and is correlated with the late time temperature fluctuations from the integrated Sachs-Wolfe (ISW) effect. 
Such a late time cross-correlation between $T$ and $y$ makes the primordial map of  
$y T$ noisy and therefore must be accounted for
in the forecasts. In the following, we model the late time $y$ distortion and its impact in enhancing the noise. We then compute the signal to noise ratio including this effect in the noise. 

In our analysis, we use the total y-parameter power spectrum of tSZ as presented in Figure 1 of 
\cite{Creque-Sarbinowski:2016wue} which is the summation of ``one-halo" and ``two-halo'' contributions in the computation of the power-spectrum (see \cite{Creque-Sarbinowski:2016wue} for more details). Using this power-spectrum, we can compute the noise associated with the y-distortion,

\ba 
\label{ClytSZ}
C_{l}^{tSZ} = C_{l}^{1h} + C_{l}^{2h},
\ea

In addition, we are also dealing with another noise associated with different experiments. Here we focus on PIXIE observation with the following instrumental noise, 
\ba
C_{l}^{\mu \mu,N} &=& w_{\mu}^{-1} \exp{\left(\frac{l^2}{l_{max}^2}\right)}, \\
C_{l}^{y y,N} &=& w_{y}^{-1} \exp{\left(\frac{l^2}{l_{max}^2}\right)} .
\ea
where $l_{max} \simeq 84$ and we have also defined $w_{\mu}^{-1} \equiv 4\pi \times 10^{-16}$ and $w_{y}^{-1} \equiv 16\pi \times 10^{-18}$.

\begin{figure}[!h]
	\centering
	\includegraphics[width=0.49\textwidth]{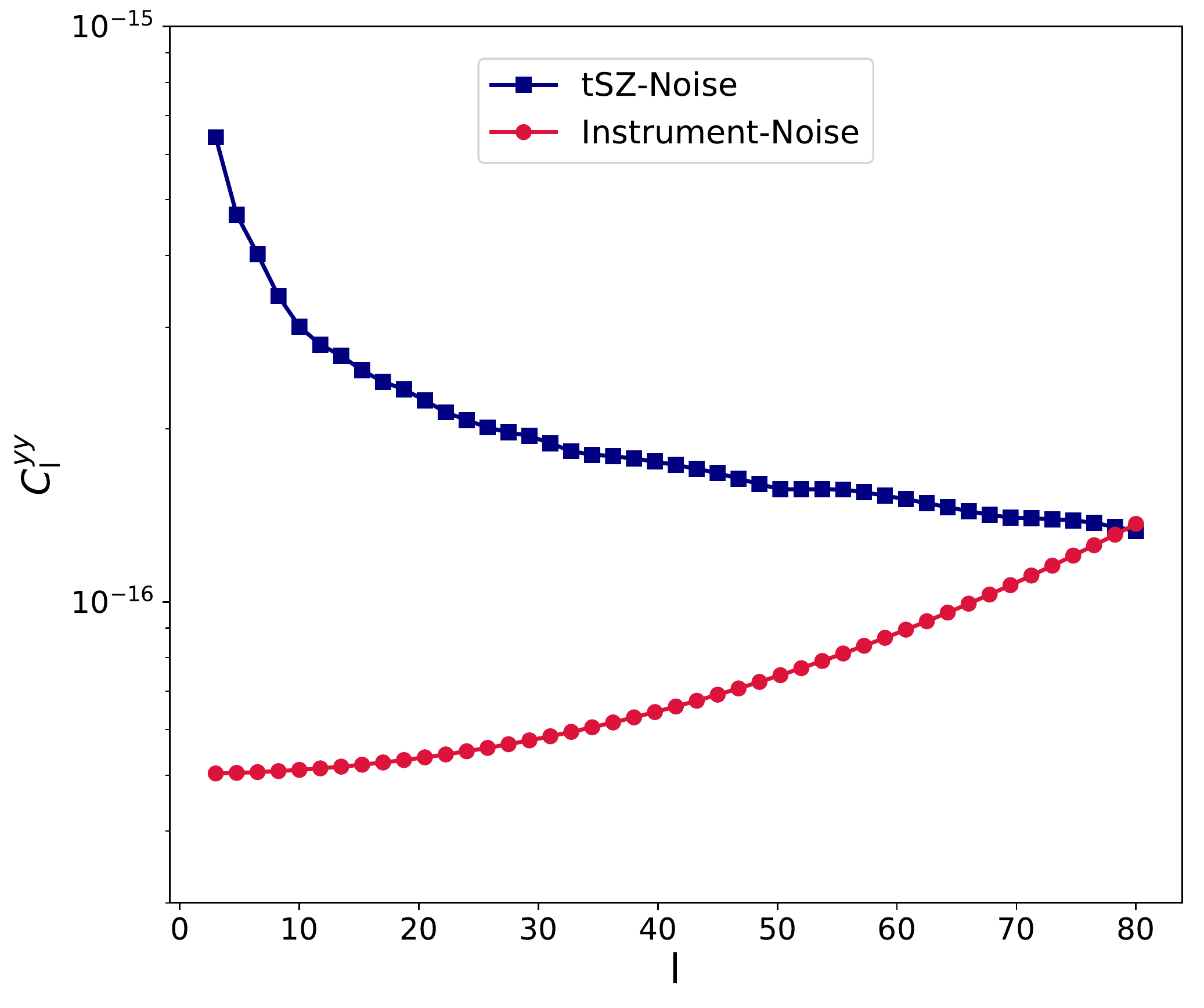}
	\hspace{0.009\textwidth}
	\caption{\label{comp} Comparison between the tSZ noise and the PIXIE instrumental noise. The plot demonstrates that tSZ effect is dominated over the instrumental noise.   }
\end{figure}

In Figure \ref{comp} we compare the tSZ noise with the instrumental noise from the PIXIE experiment. From the plot it is clear that tSZ noise is dominated over  the PIXIE instrumental noise. Owning to this, here after we almost entirely ignore the 
instrumental noise and just focus on the tSZ effect. We will however show one example in which both of the above terms are considered to demonstrate how this may affect the final constraints. 

Comparing the above results with $C_{l}^{\mu \mu}$ and $C_{l}^{yy}$, as presented in Eq. (\ref{mu-mu correlation2}), we see that
$C_{l}^{\mu \mu, N} \gg C_{l}^{\mu \mu}$ and $C_{l}^{tSZ} \gg C_{l}^{y y, N} \gg C_{l}^{y y}$. However for the temperature fluctuations we assume that the signal is much higher than the noise. This is also much higher than the late time ISW effect \cite{Creque-Sarbinowski:2016wue}. Therefore we use the CMB $TT$ correlation function ($ \frac{l(l+1)}{2 \pi} C^{TT}_{l} = \Delta^2_R/25 \simeq 10^{-10}$) in our analysis.

Having introduced different noises in the computation of the signal to noise ratio (hereafter  S/N), in the following we present the details of the analysis. 

We start with the $\mu T$ cross-correlation function. 
\ba
\label{SN muT2}
\left(\frac{S}{N}\right)_{,\mu T} &=&  \left( \sum_{l =0}^{l_{max} } \left( 2l+1 \right)  \frac{\left( C^{\mu T}_{l} \right)^2}{C^{\mu \mu, N}_{l} C^{TT}_{l} } \right) ^{1/2} \nonumber\\
&=& 4.5 \times 10^{-5} \left(\frac{ f_{NL}^p}{\Gamma_f}\right) 
 \left(\frac{k_D(z)}{\sqrt{2}k_p} \right)^{\Gamma_f} \Bigg{|}_{z_{\mu,f}}^{z_{\mu,i}} 
 \nonumber\\
&& \times
\left(\frac{\sqrt{4\pi} \times 10^{-8}}{w_{\mu}^{-1/2}}\right).
\ea
\begin{figure}[!h]
	\centering
	\includegraphics[width=0.49\textwidth]{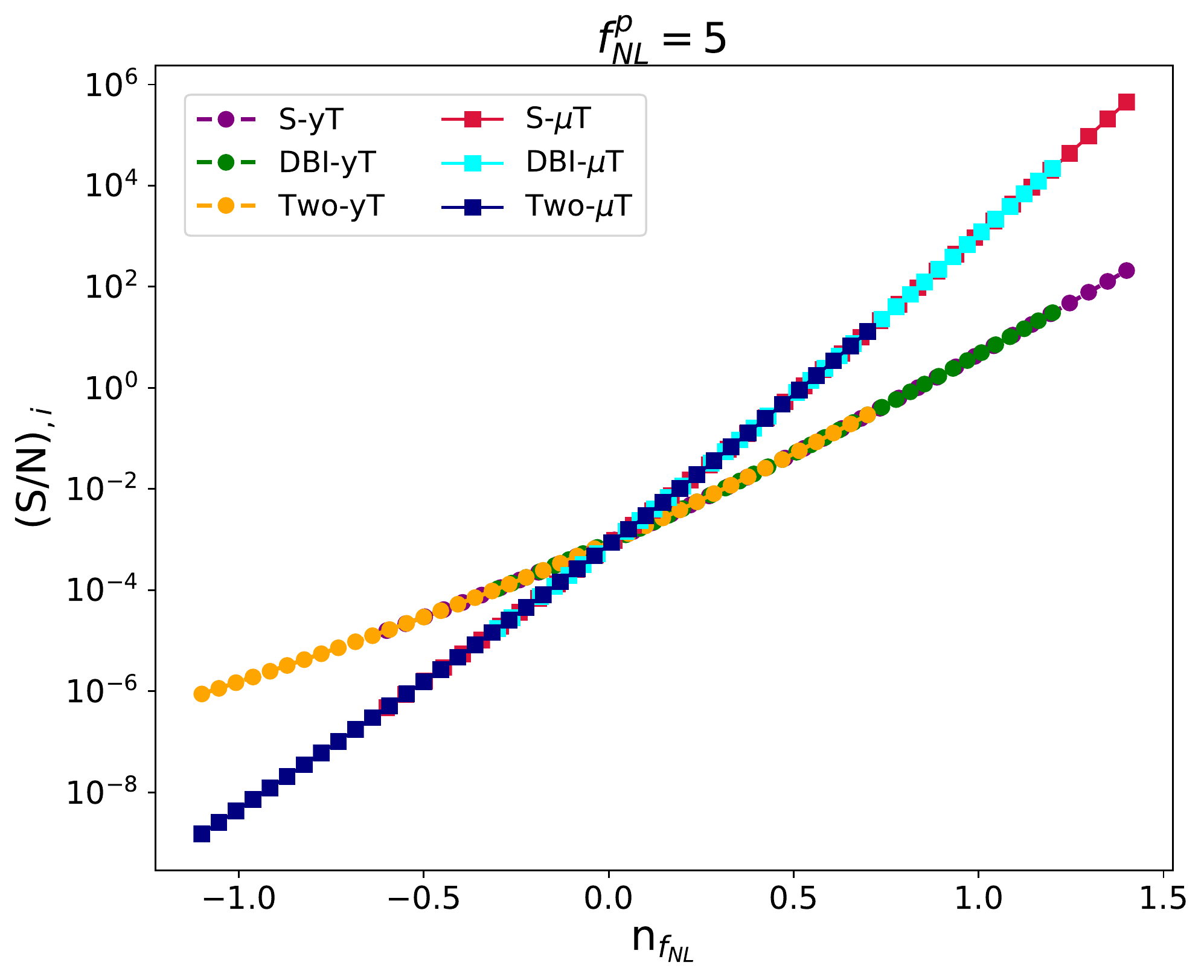}
	\hspace{0.009\textwidth}
	\caption{\label{SNdt} The behavior of $\left(S/N\right)_{,i}$ for $i = (\mu T,  yT)$ as a function of $n_{f_{NL}}$. }
\end{figure}
Next we compute the S/N for $yT$ cross-correlation function,

\ba
\label{SN yT2}
\left(\frac{S}{N}\right)_{,y T} &=&  \left(  \sum_{l =0}^{l_{max} } \left( 2l+1 \right)  \frac{\left( C^{y T}_{l} \right)^2}{C^{tSZ}_{l} C^{TT}_{l} } \right) ^{1/2} \nonumber\\
&=& \sqrt{2\pi} \frac{ f_{NL}^p \Delta^3_{\mathcal{R}}(k_p)}{\Gamma_f} \left(\frac{k_D(z)}{\sqrt{2}k_p} \right)^{\Gamma_f}\Bigg{|}_{z_{y,f}}^{z_{y,i}} \nonumber\\
&&\times 
\left(  \sum_{l =0}^{l_{max} } \frac{ \left( 2l+1 \right) }{l(l+1)} \frac{1}{(C_{l}^{yy,N} + C^{tSZ}_{l} )} \right)^{1/2}, \nonumber\\
\ea

In figure \ref{SNdt} we present S/N for Eqs. (\ref{SN muT2}) and (\ref{SN yT2}) in terms of the running of non-Gaussianity. Here we have chosen $f^p_{NL} = 5$. For $yT$ 
case, we have used both of the instrumental and tSZ effects. However, as already pointed out above, tSZ dominates over the instrumental noise.

\begin{figure*}[t!]
 \center
  \includegraphics[width=\textwidth]{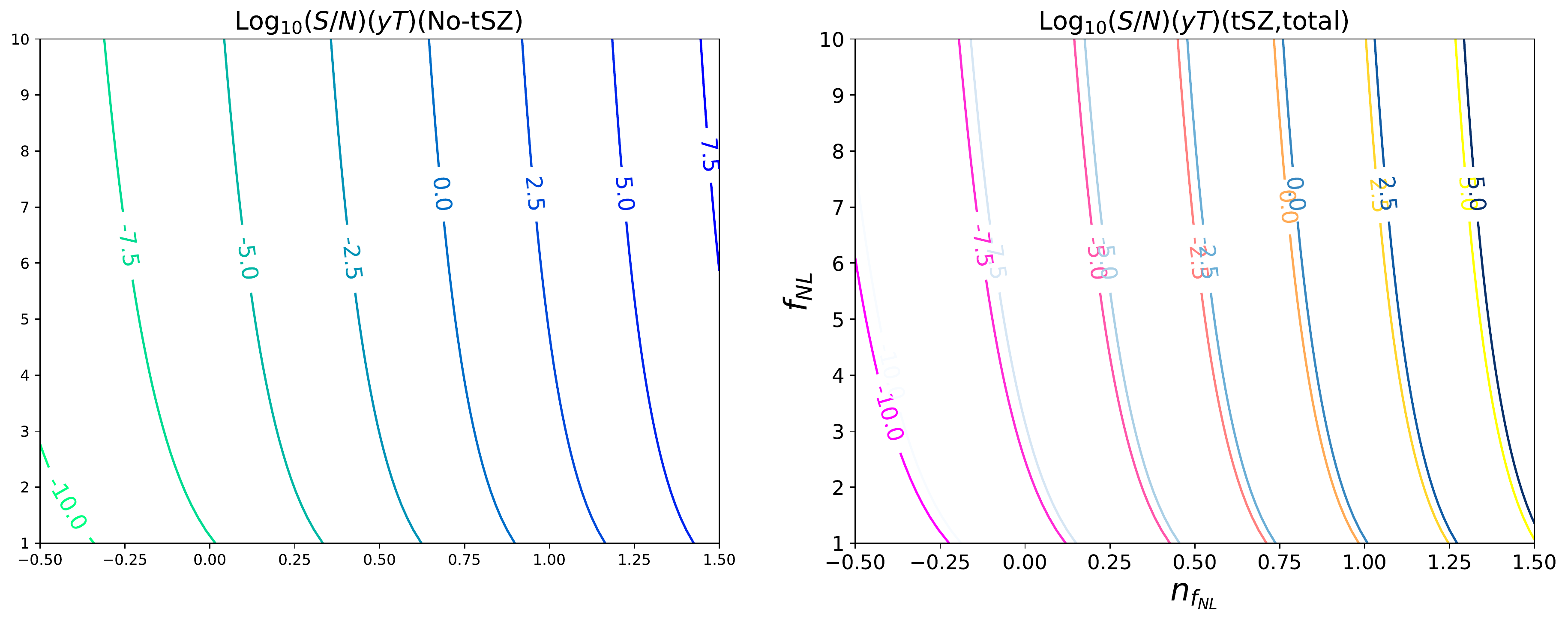}
		\caption{\label{tSZ-Inst}  Contours of $\log_{10}\left(\left( \frac{S}{N} \right)_{, yT}\right)$ including only the instrumental noise (left), tSZ effect (right:spring color map)  and both of these effects (right:blue color map ). Adding both of the instrumental and tSZ effects shift the constraints slightly to the right. }
\end{figure*}

To quantify the impact of tSZ in boosting the noise/constraints, in Figure \ref{tSZ-Inst} we estimate how much the constrains are affected by each of these noises. On the left panel, we present the impact of the instrumental noise in the signal to noise ratio, while in the right panel we consider the role of the tSZ. We consider two different cases here, one with only tSZ effect on, spring color-map, and the other with adding both of the instrumental and tSZ together in the estimation. As expected, adding tSZ effect shifts the constraints to higher values for the running of non-Gaussianity. The plot also shows that tSZ effect dominates over the instrumental noise as the plot with both of these effect on is almost exactly equal to the case with just tSZ effect. 

Although boosted due to the tSZ effect, for the first time we directly see the enhancing profile of S/N that occurs with  increasing the running of non-Gaussianity. That highlights the importance of deviating from the scale invariant non-Gaussianity as was estimated before. 

Next we compute $S/N$ for the auto/cross correlation of the SDs with each other. 

Let us begin with $\mu\mu$,
\ba
\label{SN mumu2}
\left(\frac{S}{N}\right)_{,\mu \mu} &=&   \left( \sum_{l =0}^{l_{max} } \frac{\left( 2l+1 \right)}{2}  \frac{\left( C^{\mu \mu}_{l} \right)^2}{ \left( C^{\mu \mu, N}_{l} \right) ^2} \right) ^{1/2} \nonumber\\
&=& 
 1.22 \times 10^{-10}  \left(\frac{\tau_{NL}^{p}}{\Gamma_g^2} \right)
\left(\frac{k_{D}(z)}{\sqrt{2}k_p}\right)^{2\Gamma_g}\Bigg{|}_{z_{\mu,f}}^{z_{\mu,i}}  \nonumber\\
&&  \times
\left(\frac{4\pi \times 10^{-16}}{w_{\mu}^{-1}}\right),
\ea
We continue with computing $yy$,
\ba
\label{SN yy2}
\left(\frac{S}{N}\right)_{,yy} &=& \left( \sum_{l =0}^{l_{max} } \frac{\left( 2l+1 \right)}{2}  \frac{\left( C^{yy}_{l} \right)^2}{ \left( C^{tSZ}_{l} \right) ^2} \right) ^{1/2}
\nonumber\\
&=& 
2.95 \tau^p_{NL}  \frac{ \Delta^6_{\mathcal{R}}(k_p)}{\Gamma_g^2} \left(\frac{k_{D}(z)}{\sqrt{2}k_p}\right)^{2\Gamma_g}\Bigg{|}_{z_{y,f}}^{z_{y,i}} \nonumber\\
&& \times  \left( \sum_{l =0}^{l_{max} } \frac{\left( 2l+1 \right)}{(l(l+1))^2} \frac{1}{\left( C^{tSZ}_{l} \right) ^2} \right)^{1/2}
\ea
Finally we calculate $S/N$ for $\mu y$,

\ba
\label{SN muy2}
\left(\frac{S}{N}\right)_{,\mu y}  &=&  \left( \sum_{l =0}^{l_{max} } \left( 2l+1 \right)  \frac{\left( C^{\mu y }_{l} \right)^2}{ C^{\mu \mu, N}_{l} C^{tSZ}_{l} } \right) ^{1/2} \nonumber\\
&=& 3.7 \sqrt{\pi} \tau^p_{NL} \frac{10^8 \Delta^6_{\mathcal{R}}(k_p)}{\Gamma_g^2} 
\left(\frac{k_{D}(z)}{\sqrt{2}k_p}\right)^{\Gamma_g}\Bigg{|}_{z_{\mu,f}}^{z_{\mu,i}}  
\nonumber\\
&& \times
\left(\frac{k_{D}(z)}{\sqrt{2}k_p}\right)^{\Gamma_g}\Bigg{|}_{z_{y,f}}^{z_{y,i}}  
\left(\frac{ \sqrt{4 \pi}\times 10^{-8}}{w_{\mu}^{-1/2} }\right) \nonumber\\
&& \times 
\bigg( \sum_{l =0}^{l_{max} }   \frac{\left( 2l+1 \right)}{(l(l+1))^2} 
 \exp{\left[-\frac{1}{2}\left(\frac{l}{l_{max}}\right)^2\right]} \nonumber\\
&&~~~~~ \times (C^{tSZ}_{l} )^{-1}\bigg)^{1/2}. 
\ea

\begin{figure}[h!]
	\centering
	\includegraphics[width=0.49\textwidth]{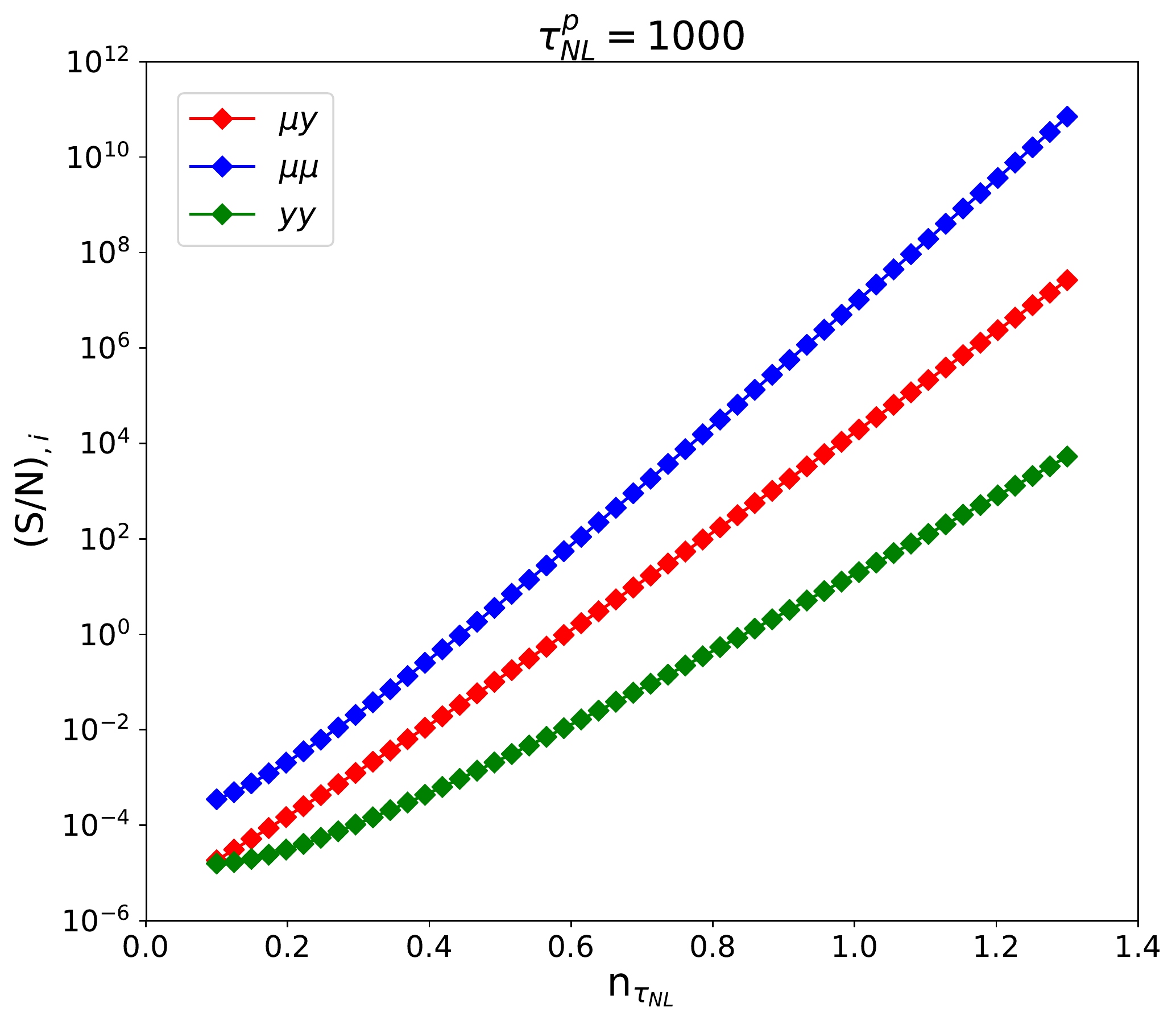}
	\caption{\label{SN-auto-correlation} The behavior of $\left(S/N\right)_{,i}$ for $i =(\mu \mu, \mu y, yy)$ as a function of $n_{\tau_{NL}}$. Here we have chosen $\tau_{NL}^{p} = 10^3$. }
\end{figure}

In figure \ref{SN-auto-correlation} we present the behavior of $S/N$ for $\mu \mu$, $\mu y$ and $yy$. Here We choose $\tau_{NL}^{p}= 10^3$. This is compatible with the recent constraints on trispectrum in \cite{Ade:2015ava}. Interestingly $S/N$ is enhances a lot by increasing the running of trispectrum. 

\section{Conclusion}
\label{conclusion}
In this paper, we calculated the impact of the running of bispectrum ($n_{f_{NL}}$) and trispectrum ($n_{\tau_{NL}}$) on the correlation functions of the CMB spectral distortion either with CMB temperature or with each other. In our analysis, we took into account the impact of production of late time y-distortion due to the thermal Sunyaev Zel'dovich effect. Assuming that any late time cross correlation between the thermal fluctuations from the integrated Sachs Wolf (ISM) and tSZ can be modeled precisely, we estimated the detectability of the primordial $yT$ correlation function in the presence of the running . 

We have shown that all of the above correlation functions are strong functions of the runnings and get enhanced by few other of magnitude upon changing the running in a reasonable range. We also computed the signal to noise ratio for all of the above correlation functions. It was shown that S/N is a sharp function of the running. It start from very small values for negative running but get enhanced by few other of magnitude for positive values. Therefore spectral distortion can possibly break the degeneracy between the positive and negative scale dependencies in the bispectrum. In addition, for moderate values of the runnings, the signal to noise ratio enhances such that makes it feasible to be detected by the 
futuristic distortion experiments like PIXIE or possibly PRISM. 

\newpage

\section*{Acknowledgment}
We thank J. Chluba, E. Dimastrogiovanni and M. Kamionkowski for their very useful discussions at the early stage of this paper and for the relevant works. We thank the anonymous referee for very constructive comments that improved the quality of this work.
R.E. acknowledges support by the Institute for Theory
and Computation at Harvard-Smithsonian Center for Astrophysics. 

\end{document}